# Indirect Microwave Holography and Through Wall Imaging

REVIEW

Okan Yurduseven[1], Michael Elsdon[2]

[1]School of Electronics, Electrical Engineering and Computer Science, Institute of Electronics, Communications and Information Technology, Queen's University Belfast, Belfast, BT3 9DT, United Kingdom
[2]Department of Mathematics, Physics and Electrical Engineering, Northumbria University, Newcastle upon Tyne, NE1 8ST, United Kingdom

Corresponding author: Okan Yurduseven, okanyurduseven@ieee.org

**ABSTRACT**
In this paper, a review of indirect microwave holography for through-wall imaging is presented. Indirect microwave holography is an imaging technique, enabling the complex object scattered fields (amplitude and phase) to be mathematically recovered from intensity-only, scalar microwave measurements. By removing the requirement to use vector measurement equipment to directly measure the complex fields, indirect microwave holography significantly reduces the cost of the imaging system and simplifies the hardware implementation. The application of a back-propagation algorithm enables the reconstructed amplitude and phase images to be obtained at the plane of the concealed object. In order to demonstrate the validity of the reviewed approach, experimental work is carried out on a metallic gun concealed under a 5 cm thick plywood wall and it is demonstrated that the indirect microwave holographic TWI can produce good resolution amplitude and phase images when back-propagation is applied. TWI of a concealed dielectric box representing non-metallic ordnance is also performed to demonstrate the ability of the technique to reconstruct through-wall images of concealed dielectric objects. An investigation of the resolution characteristics of the system suggests diffraction limited resolution can be achieved.

# 1 INTRODUCTION

Indirect microwave holography is a phase retrieval technique, enabling the complex-based object scattered fields to be mathematically retrieved from low-cost, intensity only measurements [1]. Previous work on the use of indirect microwave holography was successfully demonstrated for the measurement of complex antenna near-field and far-field radiation patterns [2-4]. Recently, this technique has been extended to the inverse scattering problem and the imaging of metallic objects has successfully been demonstrated [5-8]. Indirect microwave holography differs from conventional microwave imaging techniques in that it does not require the direct measurement of the complex field but "*indirectly*" recovers it from intensity-only measurements performed using a low-cost scalar microwave power meter. This circumvents the requirement of using expensive vector measurement equipment, thus significantly simplifying the hardware implementation and reducing the cost of the imaging system.

Through-wall imaging (TWI) of concealed objects has been the subject of much research in recent years [9-27]. Due to the advantage of microwaves to penetrate through most optically opaque materials, the use of microwaves in the detection and imaging of concealed objects is of significant importance. These applications include concealed ordnance detection, imaging of terrorist activities behind walls, archeology, detection and clearance of buried landmines, and natural disasters. In this framework, what is required is a reliable, low-cost and rapid method for the imaging of metallic and non-metallic (dielectric) concealed objects.

Solving the inverse problem is a computationally demanding task and is conventionally done by linearizing the forward-model and solving it to retrieve a qualitative estimate of the object function, such as the susceptibility distribution [28-34]. The linear inverse scattering algorithms, which are conventionally based upon a Born approximation, have the advantage of requiring a small amount of computational time due to their linear approach. However, they are limited to the determination of the geometrical features and location of objects and cannot provide information regarding the dielectric properties. On the other hand, the application of non-linear inverse scattering algorithms enables the obtainment of the dielectric properties of the concealed objects at the expense of a time-consuming reconstruction process due to their iterative approach [35-37]. Moreover, in comparison to linear inverse scattering, the non-linear approach can be unreliable due to the presence of the local minima problem. A major challenge associated to both linear and non-linear inverse scattering techniques is the requirement to use vector measurement equipment in order to measure the complex field (amplitude and phase) scattered from the imaged object. In addition to these methods, other methods demonstrated on TWI in the literature can be given as blind deconvolution [15], synthetic aperture radar (SAR) [16-19], noise radar [20, 21], multiple signal classification (MUSIC) algorithm [22], uniform geometric theory of diffraction (UTD) technique [23], compressive sensing method [24], adaptive polarization contrast technique [25], self-injection-locked (SIL) radar [26] and shifted pixel method [27].

Phase-retrieval from intensity only measurements have recently gained significant traction [38-41]. Most of this research has focused on achieving phase retrieval on the software layer by means of using iterative algorithms, such as the Wirtinger Flow algorithm studied in [38, 39]. Indirect holographic imaging enables phase retrieval from intensity only measurements without the need for an additional reconstruction algorithm on the signal processing layer. Instead, the retrieval process is embedded in the hardware layer in a holographic manner.

In this paper, we review the concept of indirect microwave holography and demonstrate its application for TWI of concealed metallic and dielectric objects. We demonstrate that, in addition to eliminating the need for direct phase measurements, the indirect microwave holography does not involve an iterative approach and no a-priori information is required in the reconstruction of the images of the concealed objects. The remainder of the paper is organized as follows. Section II provides a description of indirect microwave holographic imaging and how it is applied to TWI of concealed objects. Section III provides the experimental results taken on several objects, metallic and dielectric, while Section IV provides concluding remarks.

# 2 INDIRECT MIVROWAVE HOLOGRAPHY

**A. Theory of Indirect Microwave Holographic Imaging**

The indirect microwave holographic imaging measurement set-up is illustrated in Fig. 1(a) while a diagrammatic representation of the system is shown in Fig. 1(b).

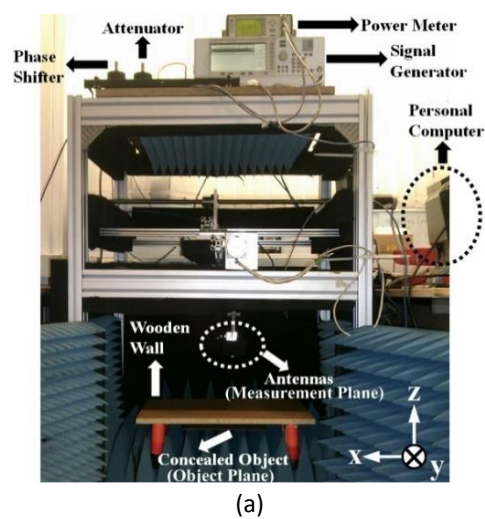

(a)

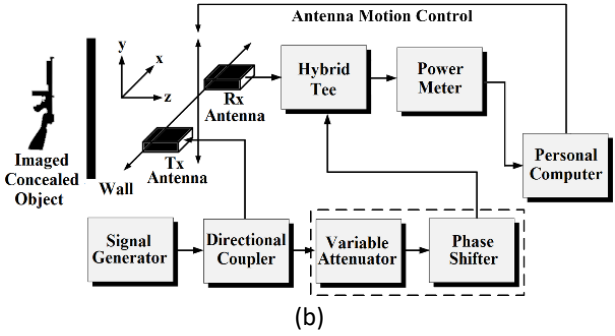

Fig. 1 Indirect microwave holographic TWI system (a) measurement set-up (b) diagrammatic representation.

As can be seen in Fig. 1, the indirect microwave holographic imaging system consists of two probe antennas, one acting as an illuminating antenna and the other one as a receiving antenna. The apertures of the probe antennas are in the measurement plane, $z=0$. A coherent reference signal generated by the RF signal generator is fed to a directional coupler where it is separated into two arms; one is fed to the illuminating antenna and the other one is connected to a block consisting of an amplitude attenuator and a phase shifter, both of which are variable. The illuminating antenna illuminates the object under imaging, which is in the object plane at a distance of $d$ from the measurement plane, $z=d$, and the scattered signal is received by the receiving antenna. The receiving antenna is connected to a hybrid tee where the received scattered signal is combined with a fraction of the reference signal tapped off through the directional coupler. The power meter, which is connected to a personal computer controlling the movement of the antennas across the scanning aperture and imports the power meter data, reads the received power consisting of the combination of the scattered signal with the tapped coherent reference signal.

Indirect microwave holographic imaging uses a two-stage process in the reconstruction of the images of scanned objects. The first stage in the imaging process is the obtainment of a holographic intensity pattern over a 2D scanning aperture at the measurement plane, $z=0$. The holographic intensity pattern of an imaged object can be given as follows:

$$I(x,y) = |E_s(x,y) + E_r(x,y)|^2 \quad (1)$$

In (1), $E_s$ is the scattered field from the imaged object while $E_r$ denotes the coherent reference wave. From (1), the following expression can be obtained

$$I(x,y) = |E_s(x,y)|^2 + |E_r(x,y)|^2 \\ + E_s^*(x,y)E_r(x,y) + E_s(x,y)E_r^*(x,y) \quad (2)$$

In (2), while the first and second components are amplitude only components, the third and fourth components are complex components and therefore consist of amplitude and phase data of the scattered field to be recovered.

The second stage in the imaging process is the processing of the holographic intensity pattern and begins with the obtainment of the Plane Wave Spectrum (PWS) of the imaged object. In order to obtain the PWS, Fourier transform of the holographic intensity pattern in (2) is taken as follows:

$$F\{I(x,y)\} = F\{|E_s(x,y)|^2\} + F\{|E_r(x,y)|^2\} \\ + F\{E_s^*(x,y)\} \otimes F\{E_r(x,y)\} + F\{E_s(x,y)\} \otimes F\{E_r^*(x,y)\} \quad (3)$$

As can be seen in (3), due to their scalar only DC content, the first and second components are not shifted and therefore they are in the center of the spectrum in the frequency domain. However, if consideration is given to the third and fourth components in (3), it can be seen that these components are convoluted by the introduced coherent reference signal and therefore shifted by an amount determined by the phase of the reference signal, which can be given as

$$E_r = \begin{cases} E_0 e^{-ik_r x}, & \text{x--axis phase shift} \\ E_0 e^{-ik_r y}, & \text{y--axis phase shift} \end{cases} \quad (4)$$

In (4), $k_r$ denotes the offset wave vector, which is given in (5) as a function of scanning sample spacing in the x-axis, $\Delta x$, and in the y-axis, $\Delta y$, respectively

$$k_r = \begin{cases} \Delta\phi/\Delta x, & \text{x--axis phase shift} \\ \Delta\phi/\Delta y, & \text{y--axis phase shift} \end{cases} \quad (5)$$

In (5), $\Delta\phi$ represents the linear phase shift between the scanning lines across the scanning aperture and can be applied in the x-axis, in the y-axis or as a combination of both. Fig. 2 demonstrates a diagram of the scanning aperture across which the linear phase shift is applied in the y-axis.

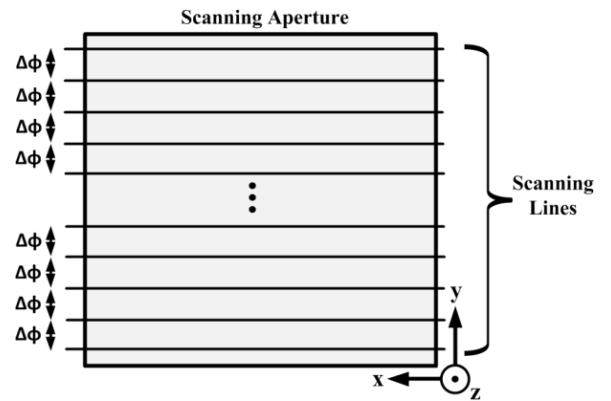

Fig. 2 Linear phase shift in the y-axis.

A diagrammatic representation of the PWS, in which the linear phase shift is applied in the y-axis, with the four components in (3) is demonstrated in Fig. 3.

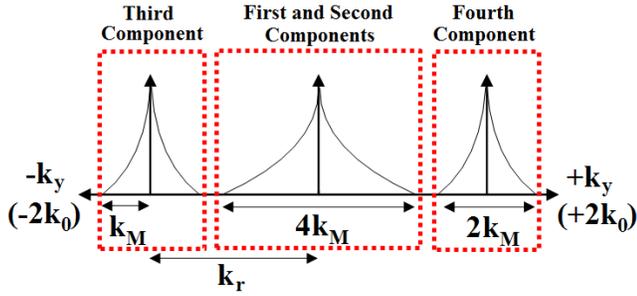

Fig. 3 Diagrammatic representation of the PWS with the four components.

The Fourier transform of the first two components in (3) are summed and placed in the center of the PWS in Fig. 3 due to their DC content. Therefore, these two components are known as central components. However, the third and fourth components highlighted in Fig. 3 are shifted towards the edges of the spectrum as a result of the convolution of the scattered signal by the introduced coherent reference signal as given in (3).

In order to be able to reconstruct the amplitude and phase images of a scanned object, possible overlaps between these components must be prevented by providing a reasonable separation between these components. Given that the PWS of the scanned object is band limited to $\pm k_M$ as demonstrated in Fig. 3, the required separation can be given as follows

$$k_r \geq 3k_M \tag{6}$$

In this paper, $\Delta\phi$ was selected as 120° and applied in the y-axis while the sample spacing in the x-axis and y-axis was 6 mm, corresponding to a quarter-wavelength in free space at the imaging frequency of 12.5 GHz, $\lambda_0/4$. This selection produces an offset wave vector of $4k_0/3$ as given in (7) where $k_0$ is the wavenumber in free space, $k_0 = 2\pi/\lambda_0$, and provides a reasonable separation between the components in the PWS.

$$k_r = \frac{\Delta\phi}{\Delta y} = \frac{(2\pi/3)}{6\,mm} = \frac{(2\pi/3)}{\lambda_0/4} = 4k_0/3 \tag{7}$$

It can be seen in (7) and Fig. 3 that introducing the offset wave vector in this manner enables the offset wave vector to exceed $k_0$ and extend into the invisible region.

From (3), it can be seen that both of the third and fourth components of the PWS in Fig. 3 contain the complex scattered field data, $E_s(x,y)$, and therefore can be used to reconstruct the images of concealed objects. In this paper, the fourth component was selected by being filtered as follows

$$F'\{I(x,y)\} = F\{E_s(x,y)\} \otimes F\{E_r^*(x,y)\} \tag{8}$$

Taking the inverse Fourier transform of (8) and reinserting the complex conjugate of the reference wave gives the original complex scattered field at the measurement plane $z=0$ multiplied by a constant as follows

$$E(x,y,z=0) = |E_r|^2 E_s(x,y) \tag{9}$$

While the amplitude of the obtained field in (9), $|E(x,y,z=0)|$, gives the reconstructed amplitude image of the imaged concealed object at the measurement plane, $z=0$, the angle data, $\angle E(x,y,z=0)$, produce the reconstructed phase image.

B. Back Propagation

In TWI of concealed objects placed at a distance of $d$ from the antennas behind the wall, significant enhancement in the reconstructed images can be achieved by using back propagation algorithms, which transform the complex scattered field at the measurement plane, $E_s(z=0)$, to the complex scattered field at the imaged object plane, $E_s(z=d)$. In order to achieve this, the back-propagation wave vector in the z-axis, $k_z$, needs to be introduced as follows

$$k_z = \begin{cases} \sqrt{k_0^2 - k_x^2 - k_y^2}, & k_0^2 \geq k_x^2 + k_y^2 \\ -i\sqrt{k_x^2 + k_y^2 - k_0^2}, & k_0^2 < k_x^2 + k_y^2 \end{cases} \tag{10}$$

In (10), $k_x$ and $k_y$ denote the propagation wave vectors in the x-axis and y-axis, respectively. If consideration is given to the filtered fourth component of the PWS in (8), the original scattered field at the object plane, $z=d$, can be obtained by taking the inverse Fourier transform of the filtered fourth component back-propagated by a distance of $d$ towards the object plane as follows

$$E'(x,y,z=d) = F^{-1}\{F'\{I(x,y)\}e^{-ik_z d}\} \tag{11}$$

Similar to the reconstruction of the amplitude and phase images at the measurement plane, $z=0$, the amplitude of the back-propagated scattered field, $|E(x,y,z=d)|$, produces the amplitude image of the object under imaging at the object plane, $z=d$, while the phase data, $\angle E(x,y,z=d)$, provide the back-propagated phase image.

3 EXPERIMENTAL RESULTS AND DISCUSSION

A. Indirect Microwave Holographic TWI of a Concealed Gun

In this Section, indirect microwave holographic TWI of a concealed metallic gun is demonstrated. The measurement set-up is illustrated in Fig. 4(a) and the concealed gun, which is made out of copper film printed upon a piece of cardboard, is demonstrated in Fig. 4(b).

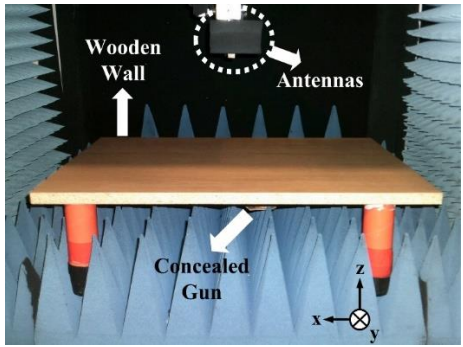
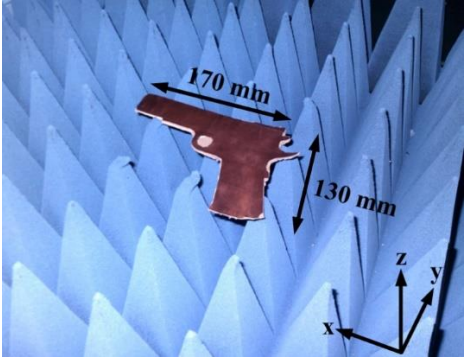

(a)

(b)

Fig. 4 Indirect holographic TWI of a concealed gun (a) experimental set-up (b) imaged concealed gun.

As can be seen in Fig 4(a), a plywood block with thickness of 5 cm in the propagation axis (z-axis) was used as a wall under which the gun was concealed at a distance of 4 cm from the block and 20 cm from the antennas in the measurement plane. The size of the scanning aperture (x-y plane) was selected as 432 mm x 432 mm with a sample spacing of 6 mm in the x-axis and y-axis corresponding to $\lambda_0/4$ at the imaging frequency of 12.5 GHz. This selection results in a square holographic intensity pattern matrix consisting of 72x72 elements.

The measured holographic intensity pattern of the concealed gun is demonstrated in Fig. 5.

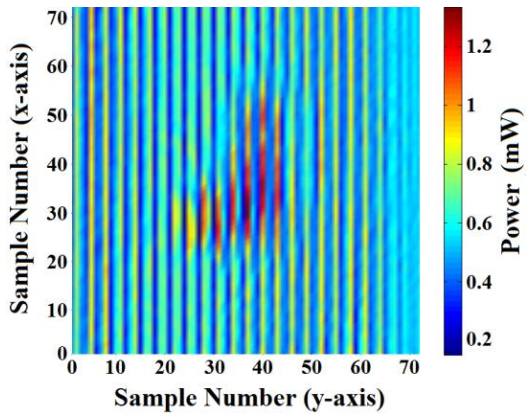

Fig. 5 Indirect holographic intensity pattern of the through-the-wall imaged concealed gun.

Following the observation of the holographic intensity pattern in Fig. 5, the intensity pattern matrix was zero padded from 72x72 to 256x256 in order to smooth the Fourier transform response of the matrix. The zero-padded 256x256 holographic intensity pattern matrix is demonstrated in Fig. 6.

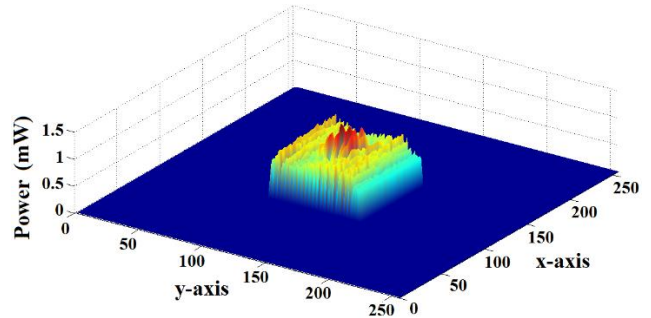

Fig. 6 Zero padded holographic intensity pattern matrix of the concealed gun.

Fourier transform of the zero padded holographic intensity pattern matrix was taken in order to obtain the PWS of the concealed gun, which is demonstrated in Fig. 7.

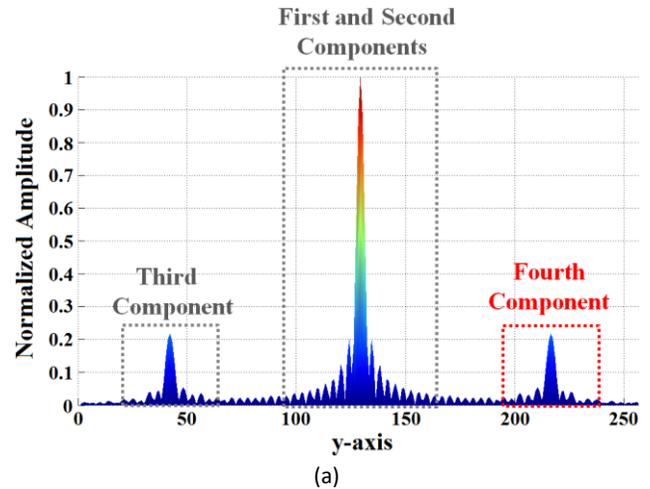

(a)

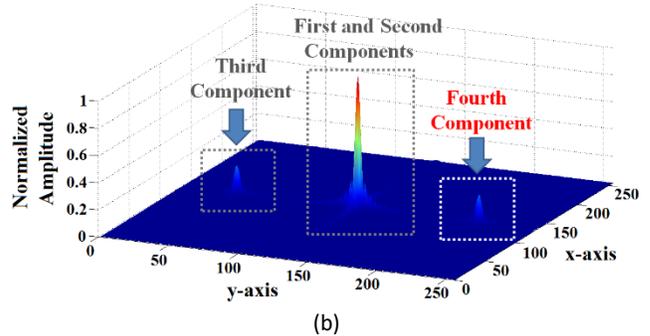

(b)

Fig. 7 PWS of the concealed gun (a) along the y-axis (b) three-dimensional

As can be seen in Fig. 7(a), a reasonable separation has been achieved between the components in the PWS and therefore no significant overlap between these components is present. As a result of applying the linear phase shift in the y-axis, separation was achieved in the y-axis while no separation is present in the x-axis as demonstrated in Fig. 7(b). In the PWS of the concealed gun demonstrated in Fig. 7, both the third and fourth components close to the edges of the spectrum include the required complex scattered field data from the gun. In this paper, the fourth component highlighted in Fig.

7(a) was used to reconstruct the scattered complex field while the central and third components were filtered off.

For applications where linear phase shift and sample spacing selections do not provide enough separation between the PWS components to achieve a proper component filtering due to an overlap, two solutions can be applied. One way to overcome this challenge is to decrease the sample spacing to increase the offset wave vector $k_r$ as given in (5) which would also result in an increase in the number of the sampling points required for the imaging. Therefore, this solution would increase the required total measurement and computational time. Another way is to remove the central DC component of the PWS by subtracting the average value of the holographic intensity pattern matrix from the original matrix prior to taking the Fourier transform as given below

$$I(x, y) = I(x, y) - \text{avarage}(I(x, y)) \quad (12)$$

In this case the PWS spectrum of the concealed gun is illustrated in Fig. 8.

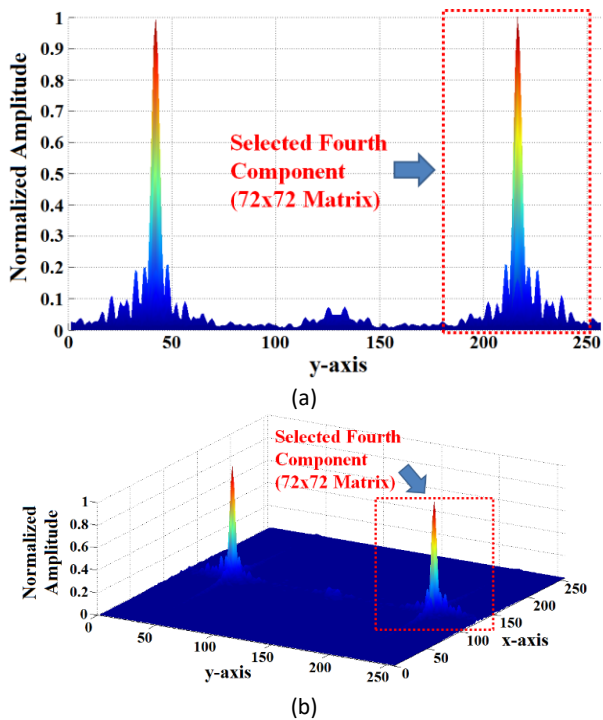

Fig. 8 PWS of the concealed gun after the average value of the intensity matrix was subtracted prior to being Fourier transformed (a) along the y-axis (b) three-dimensional.

As illustrated in Fig. 8, as a result of subtracting the average value of the holographic intensity pattern matrix, the central component has successfully been removed, resulting in a greater separation between the components of the PWS.

In the selection of the fourth PWS component in Fig. 8, consideration must be given to determine the optimum filtering size. This is because while the sidelobes close to the main lobe of the filtered fourth component contain low-frequency data determining the general structure in the reconstructed images, other lobes further away from the main lobe contain high-frequency data determining the small details such as the sharpness of the edges in the reconstructed images. Therefore, improperly selecting the filtering size would significantly deteriorate the image reconstruction accuracy of the proposed technique. To this end, parametric analysis was carried out and a filtering size of 72x72 was found to be an ideal PWS fourth component filtering matrix size to achieve optimum image reconstruction accuracy. Following the selection of the fourth PWS component, inverse Fourier transform was applied to recover the complex scattered field from which the amplitude and phase images of the concealed gun at the measurement plane, $z=0$, are reconstructed as shown in Fig. 9 with the actual outline of the gun highlighted.

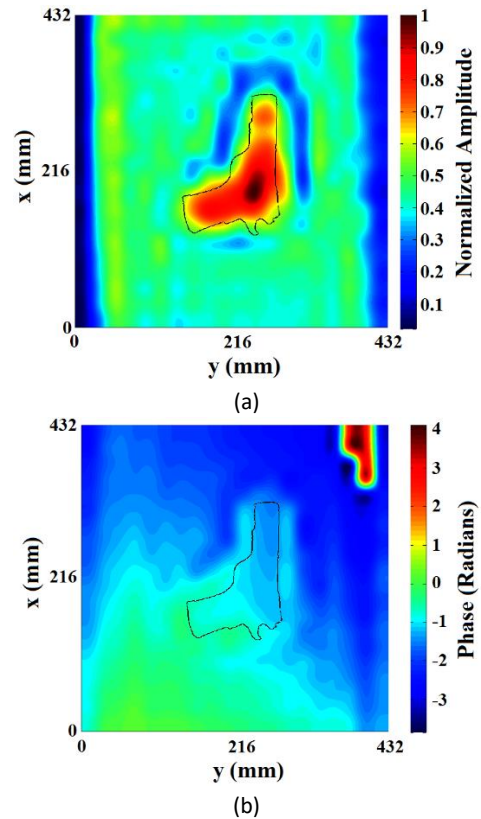

Fig. 9 Reconstructed through-the-wall images of the concealed gun at the measurement plane, $z=0$ (a) reconstructed amplitude (b) reconstructed phase.

As can be seen in Fig. 9(a), at the measurement plane $z=0$, while the reconstructed amplitude image provides a reasonable contrast revealing the approximate shape of the imaged concealed gun, the reconstructed phase image in Fig. 9(b) does not provide any useful information in the TWI detection and identification of the gun. Therefore, further improvement is required, which can be achieved by applying the back-propagation algorithm described in Section II-B. In Fig. 10, the reconstructed amplitude and phase images of the concealed gun back-propagated to the object plane behind the wall, $z=200$ mm, are demonstrated.

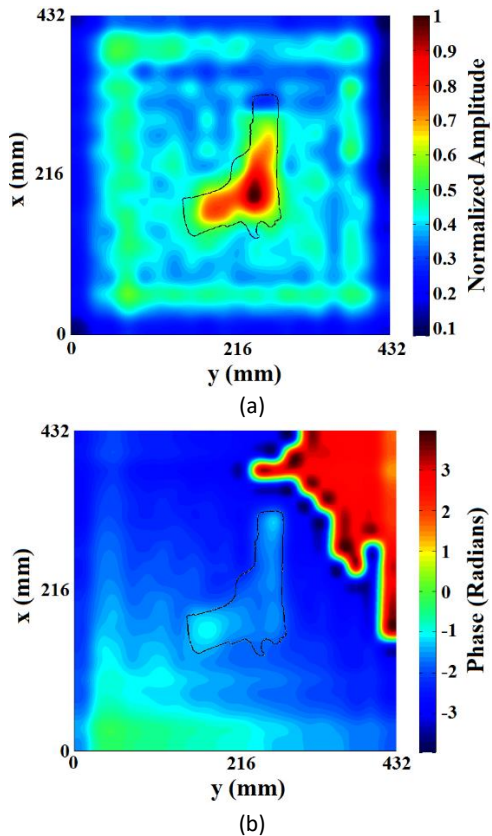

(a)

(b)

Fig. 10 Reconstructed back-propagated through-the-wall images of the concealed gun at the object plane, $z$=20 cm (a) reconstructed amplitude (b) reconstructed phase.

If comparison needs to be made between the reconstructed images at the measurement plane in Fig. 9 and at the object plane in Fig. 10, the improvement in the approximation of the reconstructed images to the actual shape of the concealed gun is evident. Particularly, in Fig. 10(b), the reconstructed phase image at the object plane is in good agreement with the actual outline and position of the imaged gun.

**B. Indirect Microwave Holographic TWI of a Concealed Box**

In this Section, indirect holographic TWI of a dielectric box representing dielectric ordnance is performed in order to demonstrate the ability of the proposed technique to image not only concealed metallic objects but also concealed dielectric objects. The TWI measurement set-up of the concealed dielectric box is illustrated in Fig. 11.

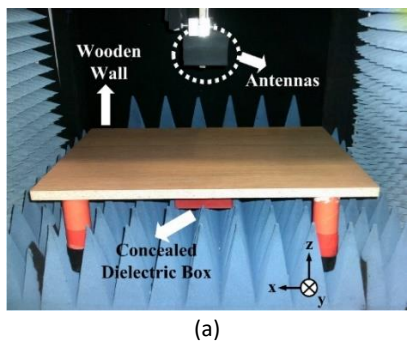

(a)

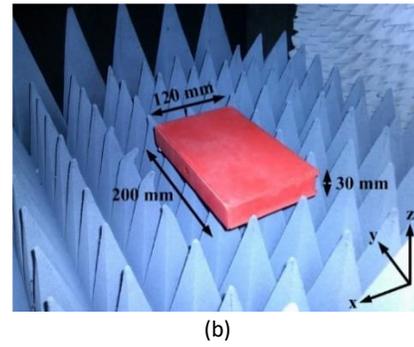

(b)

Fig. 11 Indirect holographic TWI of a dielectric box representing a plastic explosive (a) experimental set-up (b) imaged dielectric box.

As can be seen in Fig. 11, the rectangular box was concealed under the wall at a distance of 2 cm while the separation between the antennas in the measurement plane and the box was 18 cm. The measurement was taken over a scanning aperture of 432 mm x 432 mm with a sample spacing of 6 mm in the x-axis and y-axis resulting in a 72x72 element square holographic intensity pattern matrix. The measured holographic intensity pattern is demonstrated in Fig. 12.

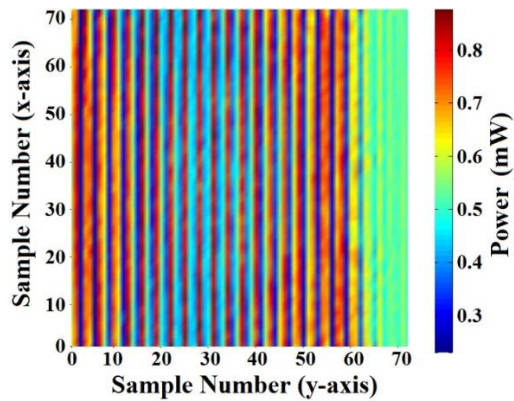

Fig. 12 Indirect holographic intensity pattern of the through-the-wall imaged dielectric box.

Fourier transform of the 72x72 holographic intensity pattern matrix demonstrated in Fig. 12 was taken after the matrix was zero padded to 256x256 and the average value of the matrix was subtracted to remove the central DC PWS component. The obtained PWS of the concealed dielectric box is demonstrated in Fig. 13.

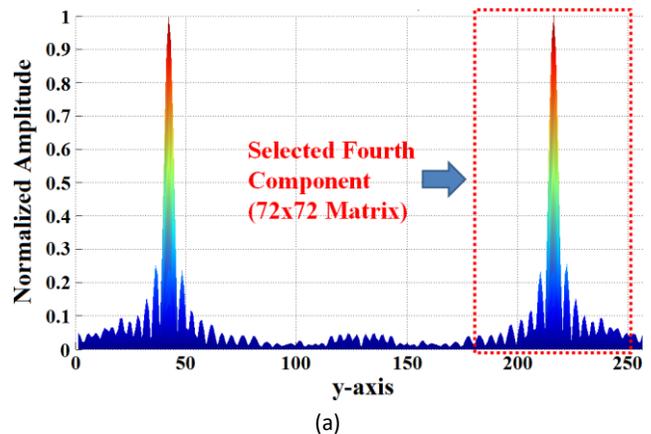

(a)

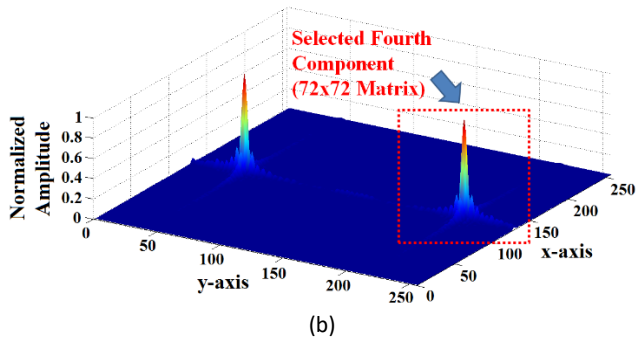

(b)

Fig. 13 PWS of the concealed dielectric box after the average value of the intensity matrix was subtracted prior to being Fourier transformed (a) along the y-axis (b) three-dimensional.

As can be seen in Fig. 13, good separation between the PWS components was achieved as a result of applying a linear phase shift of 120° in the y-axis and selecting the sample spacing as 6 mm. In the PWS in Fig. 13(a), the highlighted fourth component was selected and inverse Fourier transformed in order to recover the complex scattered field at the measurement plane, $z=0$, from which the magnitude and phase images of the concealed dielectric box were reconstructed as illustrated in Fig. 14. In Fig. 14, the dashed rectangle denotes the position and extent of the original object.

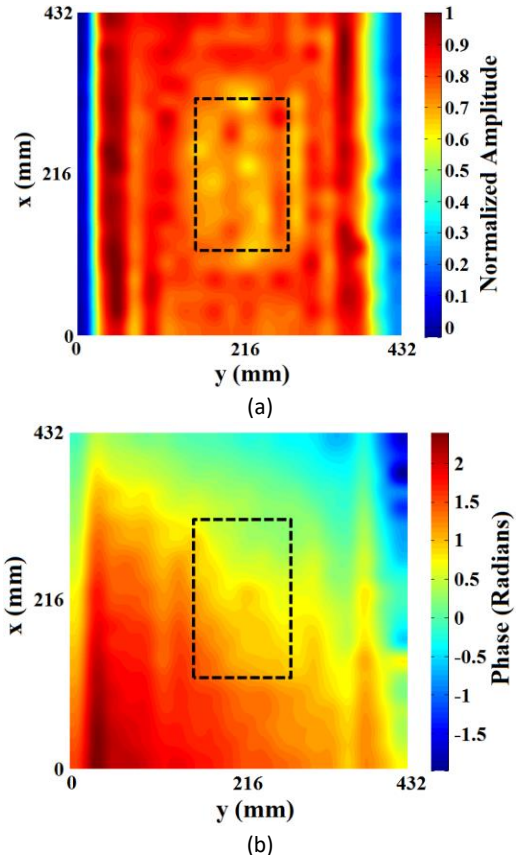

Fig. 14 Reconstructed through-the-wall images of the concealed dielectric box at the measurement plane, $z=0$ (a) reconstructed amplitude (b) reconstructed phase.

As demonstrated in Fig. 14(a), although the reconstructed magnitude image at the measurement plane reveals the presence of an object concealed under the wall, it is rather difficult to discern the actual shape of the box. From the reconstructed phase image at the measurement plane demonstrated in Fig. 14(b), on the other hand, no information can be obtained regarding the shape and presence of the concealed box and therefore back-propagation needs to be applied in order to enhance the reconstructed images at the measurement plane. In order to obtain the complex scattered field at the plane of the concealed box, $z=18$ cm, back-propagation was applied to the selected fourth PWS component followed by an inverse Fourier transform. The reconstructed amplitude and phase images of the dielectric box at the object plane are demonstrated in Fig. 15.

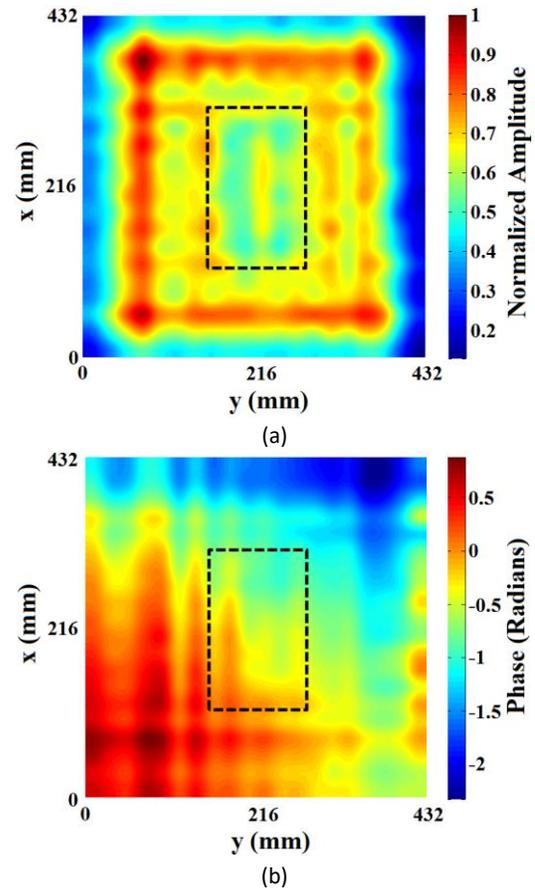

Fig. 15 Reconstructed back-propagated through-the-wall images of the concealed dielectric box at the object plane, $z=18$ cm (a) reconstructed amplitude (b) reconstructed phase.

As can be seen in Fig. 15(a), the reconstructed magnitude image at the object plane clearly reveals the presence of the dielectric rectangular box concealed under the wall while the back-propagated reconstructed phase image in Fig. 15(b) provides an outline of the box.

## C Indirect Microwave Holographic TWI of Concealed Coins

In order to demonstrate the resolution limits of the proposed indirect microwave holographic TWI technique, the imaging of two 5p British coins concealed under the wooden wall was carried out. The coins have a diameter of 18 mm corresponding to a size of smaller than $\lambda_0$ at the imaging frequency of 12.5 GHz. The measurement set-up is demonstrated in Fig. 16.

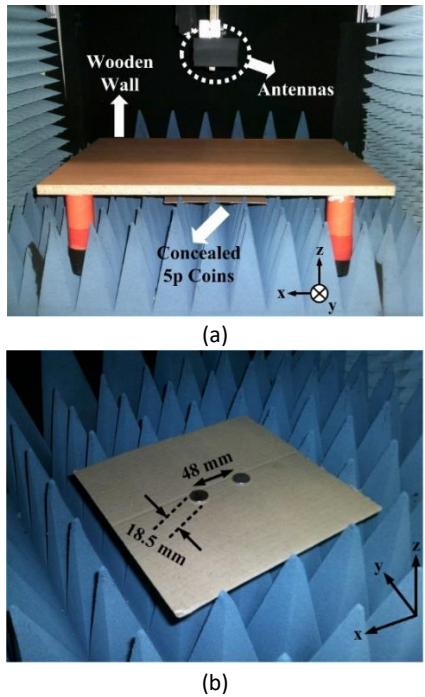

(a)

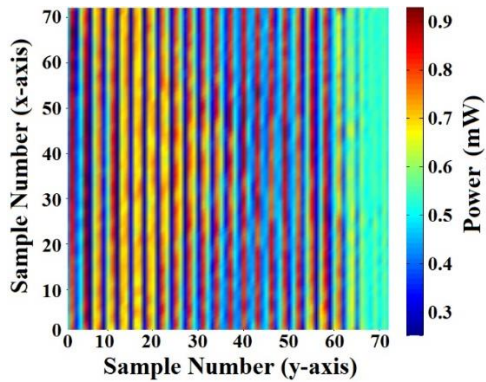

(b)

Fig. 16 Indirect holographic TWI of two 5p British coins (a) experimental set-up (b) imaged coins.

As demonstrated in Fig. 16, the coins were separated by a distance of 48 mm from each other, which corresponds to $2\lambda_0$ at the frequency of 12.5 GHz, and placed upon a piece of cardboard. The coins were concealed under the wooden wall at a distance of $d$=4 cm from the wall and $d$=20 cm from the antennas in the measurement plane. The obtained holographic intensity pattern is demonstrated in Fig. 17.

Fig. 17 Indirect holographic intensity pattern of the through-the-wall imaged concealed 5p British coins.

Following the observation of the holographic intensity pattern matrix, it was zero-padded to 256x256 and underwent an average value subtraction prior to being Fourier transformed. Fourier transform of the zero-padded intensity pattern matrix was taken as described in (3) in order to obtain the PWS of the concealed coins, which is demonstrated in Fig. 18.

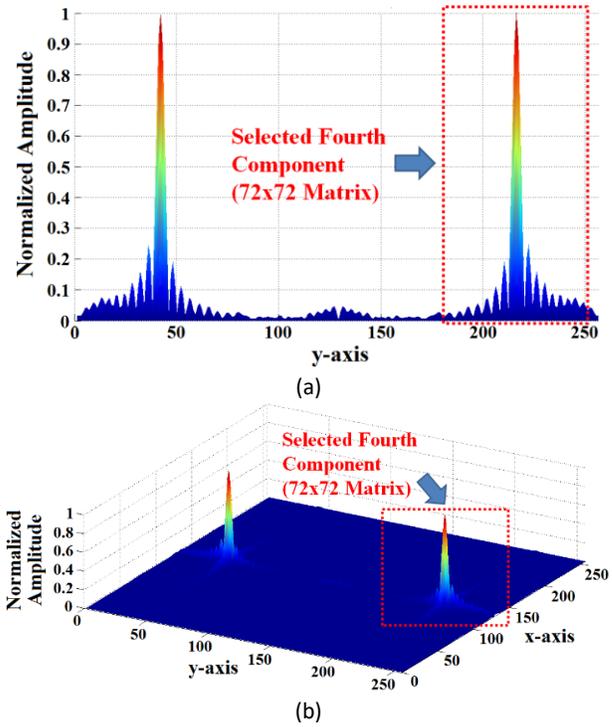

Fig. 18 PWS of the concealed coins after the average value of the intensity matrix was subtracted prior to being Fourier transformed (a) along the y-axis (b) three-dimensional.

As can be seen in Fig. 18, as a result of subtracting the average value of the holographic intensity pattern matrix, the DC central component of the PWS has successfully been filtered off. The selection of the linear phase shift as 120° and sample spacing as 6 mm, on the other hand, resulted in a reasonable separation between the PWS components. As both the third and fourth components close to the edges of the spectrum in Fig. 18 include the required complex field scattered from the coins, both of them can be used in order to obtain the reconstructed amplitude and phase images. In a similar manner to the imaging of the concealed gun and dielectric box, the fourth component highlighted in Fig. 18 was used for the reconstruction of the amplitude and phase images of the concealed coins and therefore the third component of the PWS was filtered off.

The reconstructed amplitude and phase images of the coins at the measurement plane, $z$=0, are demonstrated in Fig. 19. In Fig. 19, the original position and outline of the imaged coins have been shown using dashed circles. As can be seen in Fig. 19, from the reconstructed images at the measurement plane, $z$=0, it is rather difficult to discern a clear outline of the coins due to their considerably small size and concealed position.

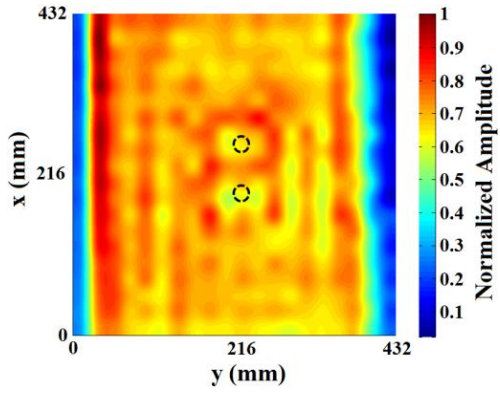

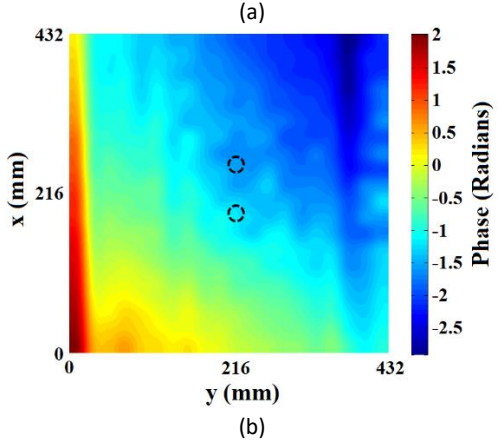

Fig. 19 Reconstructed through-the-wall images of the coins at the measurement plane, z=0 (a) reconstructed amplitude (b) reconstructed phase.

In order to enhance the imaging resolution of the coins in the reconstructed images, back-propagation was applied and the back-propagated reconstructed amplitude and phase images of the coins at the object plane, z=20 cm, are demonstrated in Fig. 20.

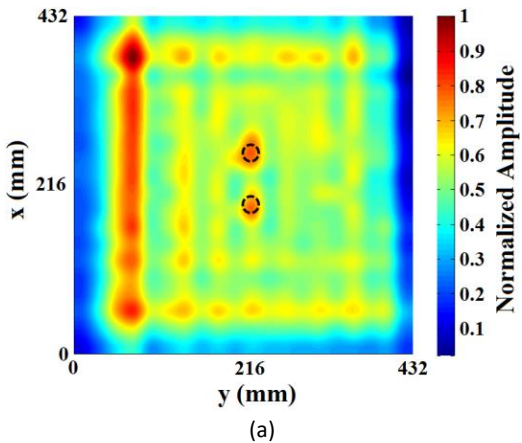

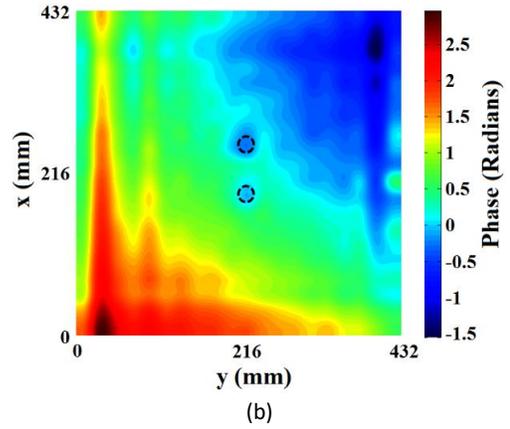

Fig. 20 Reconstructed back-propagated through-the-wall images of the coins at the object plane, z=20 cm (a) reconstructed amplitude (b) reconstructed phase.

The enhancement in the back-propagated reconstructed images of the concealed coins at the object plane in Fig. 20 is evident in comparison to the reconstructed images of the coins at the measurement plane in Fig. 19. While the back-propagated reconstructed amplitude image in Fig. 20(a) clearly reveals the presence of the coins concealed under the wall, the back-propagated reconstructed phase image in Fig. 20(b) provides a clear outline of the concealed 5p coins significantly assisting the identification of the geometric features of the concealed object. The diffraction limited resolution of the synthesized aperture can be calculated as follows:

$$\delta = \frac{\lambda_0 d}{L} \quad (13)$$

In (13), $L$ is the size of the synthesized aperture $L$=432 mm. At the operating frequency, 12.5 GHz, the theoretical resolution limit is calculated to be around 12 mm. Analyzing the reconstructed amplitude and phase images in Fig. 20, we can clearly distinguish the coins of 18 mm diameter, suggesting that the resolution of the system is diffraction limited.

## 4 CONCLUSIONS

This paper has demonstrated the use of indirect microwave holography for TWI of concealed metallic and dielectric objects. It has shown how the complex scattered field from the object under imaging can be mathematically recovered from low-cost intensity-only scalar microwave measurements. This feature differentiates indirect microwave holographic imaging from conventional inverse scattering and microwave tomography imaging methods requiring the use of expensive vector measurement equipment to directly measure the complex scattered field (amplitude and phase). The reconstructed amplitude and phase images have been demonstrated for a number of through-wall concealed objects, including a printed metallic gun, a dielectric rectangular box and two coins, and back-propagated results have demonstrated the ability of the proposed method to recover good quality images with a diffraction limited resolution. The proposed technique has the potential to be employed in a large number of imaging applications, including airport security imaging systems, concealed weapons detection and non-destructive testing.